\documentclass[aps,prd,reprint,balancelastpage,nofootinbib,preprintnumbers,superscriptaddress,onecolumn,notitlepage]{revtex4-1}

\usepackage{slashed}
\usepackage{bm}
\usepackage{latexsym,amssymb,amsmath,float,url,mathrsfs}
\usepackage{latexsym}
\usepackage{graphicx}
\usepackage{epstopdf}
\usepackage{amsmath}
\usepackage{comment}
\usepackage{subfigure}
\usepackage{natbib}
\usepackage{hyperref}
\usepackage{pifont}
\usepackage{blindtext}
\usepackage{adjustbox}
\usepackage{multirow}
\usepackage{tabularx}
\usepackage[table]{xcolor}
\usepackage{orcidlink}
\usepackage{tikz}
\usetikzlibrary{tikzmark}
\usepackage{fontawesome}
\usepackage{mdframed}

\usepackage{blkarray}
\usepackage{graphicx}
\usepackage{amsfonts}
\usepackage{soul}
\usepackage{amssymb}
\usepackage{amsmath}
\usepackage{cancel}
\usepackage{tcolorbox}

\usepackage{graphics,appendix,afterpage,makecell} 

\def\dd{{\rm d}}

\definecolor{oucrimsonred}{rgb}{0.6, 0.0, 0.0}
\definecolor{persianblue}{rgb}{0.11, 0.22, 0.73}
\definecolor{forestgreen}{rgb}{0.13,0.35,0.13}
\definecolor{lightgray}{rgb}{0.83, 0.83, 0.83}
 \hypersetup{colorlinks, citecolor=oucrimsonred, linkcolor=black, urlcolor=oucrimsonred}
\definecolor{cornellred}{rgb}{0.7, 0.11, 0.11}
\definecolor{navyblue}{rgb}{0.0, 0.0, 0.5}
\definecolor{amethyst}{rgb}{0.6, 0.4, 0.8}
\definecolor{yellow}{rgb}{1.0, 1.0, 0.0}
\definecolor{firebrick}{rgb}{0.7, 0.13, 0.13}
\definecolor{tangerineyellow}{rgb}{1.0, 0.8, 0.0}
\definecolor{deepfuchsia}{rgb}{0.76, 0.33, 0.76}
\definecolor{amber}{rgb}{1.0, 0.75, 0.0}
\definecolor{VioletRed4}{rgb}{0.55, 0.13, .32}
\definecolor{indiagreen}{rgb}{0.07, 0.53, 0.03}
\definecolor{VioletRed4}{rgb}{0.55, 0.13, .32}
\newcommand{\be}{\begin{equation}}
\newcommand{\ee}{\end{equation}}
\newcommand{\bea}{\begin{equation} \begin{aligned}}
\newcommand{\eea}{\end{aligned} \end{equation}}

\definecolor{oucrimsonred}{rgb}{0.6, 0.0, 0.0}
  
\newcommand{\x}{{\bf x}}  
\newcommand{\g}{{\rm g}}

\newcommand{\orm}{\overline{r}_m}

\newcommand{\td}{{\rm d}}

\newcommand{\arXiv}[2]{\href{http://arxiv.org/pdf/#1}{{\tt [#2/#1]}}}
\newcommand{\arXivold}[1]{\href{http://arxiv.org/pdf/#1}{{\tt [#1]}}}
\newcommand{\prljournal}[1]{\href{https://journals.aps.org/prl/abstract/10.1103/#1}{{\tt [#1]}}}
\newcommand\vertarrowbox[3][6ex]{%
  \begin{array}[t]{@{}c@{}} #2 \\
  \left\uparrow\vcenter{\hrule height #1}\right.\kern-\nulldelimiterspace\\
  \makebox[0pt]{\scriptsize#3}
  \end{array}%
}

\definecolor{verdechiaro}{rgb}{0.6,1,0.6}
\definecolor{giallochiaro}{rgb}{1,1,0.6}
\definecolor{bluscuro}{rgb}{0.15, 0.2, 0.9}
\definecolor{verdes}{rgb}{0.1, 0.5, 0.1}%
\definecolor{tangerineyellow}{rgb}{1.0, 0.8, 0.0}

\definecolor{americanrose}{rgb}{1.0, 0.01, 0.24}
\definecolor{cobalt}{rgb}{0.0, 0.28, 0.67}
\definecolor{brandeisblue}{rgb}{0.0, 0.44, 1.0}
\definecolor{mycolor}{rgb}{0.0, 0.0, 0.5}
\definecolor{oxfordblue}{rgb}{0.0, 0.13, 0.28}
\definecolor{azure}{rgb}{0.0, 0.5, 1.0}
\definecolor{turquoiseblue}{rgb}{0.0, 1.0, 0.94}
\newtcolorbox{mynewbox}[1]{colback=white!5!white,colframe=azure!75!black,fonttitle=\bfseries,title=#1}
\newtcolorbox{mybox}{colback=mycolor!5!white,colframe=azure!75!black}
\newtcolorbox{mynamedbox}[1]{colback=mycolor!5!white,colframe=azure!75!black,title=#1}
\definecolor{venetianred}{rgb}{0.78, 0.03, 0.08}
\newtcolorbox{mynamedbox1}[1]{colback=venetianred!5!white,colframe=venetianred!80!black,title=#1}
\newtcolorbox{mynamedbox2}[1]{colback=azure!5!white,colframe=azure!80!black,title=#1}

\definecolor{verdes}{rgb}{0.1, 0.5, 0.1}%
\definecolor{cornellred}{rgb}{0.7, 0.11, 0.11}

\definecolor{VioletRed4}{rgb}{0.55, 0.13, .32}

\hypersetup{
     colorlinks   = true,
     citecolor    = violet,
     urlcolor     = violet,
     linkcolor    = violet}

\definecolor{rossocorsa}{rgb}{0.83, 0.0, 0.0}

\usepackage[normalem]{ulem}

\newcommand{\papertitle}{The Primordial Black Hole Abundance:
The Broader, the Better}

\begin{document}

\title[]{\papertitle}
\author{Andrea Ianniccari\orcidlink{0009-0008-9885-7737}}
\affiliation{Department of Theoretical Physics and Gravitational Wave Science Center,  \\
24 quai E. Ansermet, CH-1211 Geneva 4, Switzerland}

\author{Antonio J. Iovino\orcidlink{0000-0002-8531-5962}}
\affiliation{Dipartimento di Fisica, ``Sapienza'' Universit\`a di Roma, Piazzale Aldo Moro 5, 00185, Roma, Italy}
\affiliation{INFN Sezione di Roma, Piazzale Aldo Moro 5, 00185, Roma, Italy}

\author{Alex Kehagias\orcidlink{0000-0001-6080-6215}}
\affiliation{Physics Division, National Technical University of Athens, Athens, 15780, Greece}

\author{Davide Perrone\orcidlink{0000-0003-4430-4914}}
\affiliation{Department of Theoretical Physics and Gravitational Wave Science Center,  \\
24 quai E. Ansermet, CH-1211 Geneva 4, Switzerland}

\author{Antonio Riotto\orcidlink{0000-0001-6948-0856}}
\affiliation{Department of Theoretical Physics and Gravitational Wave Science Center,  \\
24 quai E. Ansermet, CH-1211 Geneva 4, Switzerland}


\begin{abstract}
\noindent
We show that the abundance of primordial black holes, if formed through the collapse of large fluctuations generated during inflation and unless the power spectrum of the curvature perturbation is very peaked,  is always dominated by the broadest profile of the compaction function, 
where the corresponding threshold is 2/5, even though statistically it is not the most frequent. This result exacerbates the tension when combining the primordial black hole abundance with  the signal seen by pulsar timing arrays  and  originated from gravitational waves  induced by the same large primordial perturbations.   
 
\end{abstract}

\maketitle
\vspace{1cm}
\section{Introduction and Conclusions}  \label{sec:intro}
\noindent
Primordial Black Holes (PBHs) have emerged as one of the most interesting topics in cosmology in the last years (see Ref. \cite{LISACosmologyWorkingGroup:2023njw} for a recent review).   PBHs could explain both some of the signals from binary black hole mergers observed in gravitational wave detectors \cite{Franciolini:2021tla} and be an important component of the dark matter in the Universe. 

One of the crucial parameter in PBHs physics is the relative abundance of PBHs with respect to the dark matter component. This quantity is not easy to calculate in the scenario in which PBHs are formed by the collapse of large fluctuations generated during inflation upon horizon re-entry. Indeed, 
the  formation probability is very sensitive to tiny changes in the various ingredients, such as  the critical threshold of collapse, the non-Gaussian nature of the fluctuations,  the choice of the window function to define smoothed observables (see again Ref. \cite{LISACosmologyWorkingGroup:2023njw} for a nice discussion on such issues), the  nonlinear corrections entering in the calculation of the PBHs abundance from  the 
 nonlinear radiation transfer function and 
the determination of the true physical horizon crossing  \cite{DeLuca:2023tun} and the  appearance of an infinite tower of local, non-local and  higher-derivative operators upon dealing with the nonlinear overdensity \cite{Franciolini:2023wun}.

One intrinsic and therefore unavoidable source of uncertainty in calculating the PBHs abundance arises from the inability to predict the value of a given observable with zero uncertainty, e.g. the compaction function or its  curvature at its peak, in a given point or region. This is due to the fact that the theory delivers only stochastic quantities, e.g. the curvature perturbation, of which we know only the power spectrum and the  higher-order correlators. Therefore, we are allowed    to calculate only ensemble averages and  typical values, which come with  intrinsic uncertainties quantified by, for example,  root mean square deviations. 

Since the critical PBHs abundance depends crucially on the curvature of the compaction function at its peak, the natural question which arises is the following: in order to calculate the PBHs abundance, which value of the critical threshold should we use? In other words, which value of the curvature should one adopt to derive the formation threshold?

A natural answer to this question might be to use the average profile of the compaction function, and this is done routinely in the literature. After all, most of the Hubble volumes are populated by peaks with such average profile at horizon re-entry.

In this  paper we wish to make a  simple, but  relevant observation:
only if the power spectrum of the curvature perturbation is very peaked, the critical threshold for formation is determined by the average value of the curvature of the compaction function at the peak; in the realistic cases in which   the power spectrum of the curvature perturbation is not peaked,  the critical threshold for formation is determined by the broadest possible compaction function. This is because the  abundance is dominated by the smallest critical threshold, which corresponds to the broadest profile. In such a case, the threshold for the compaction function is fixed to be 2/5.

This paper is organized as follows. In section 2 we briefly summarize the properties of the compaction function; in sections  3 and 4  we prove our observation, while in section 5 we make a comparison with the recent literature and its implication with Pulsar Timing Arrays experiments. In section 6 we provide some final comments.

\section{The compaction function}\label{sec1}
\setcounter{section}{2}
\noindent
The key starting  object  is the  curvature perturbation $\zeta(\bf x)$ on superhorizon scales which appears in the metric in the comoving uniform-energy density gauge 
\be
{\rm d}s^2=-{\rm d}t^2+a^2(t)e^{2\zeta(\bf x)}{\rm d}{\bf x}^2,
\ee
where  $a(t)$ is the scale factor in terms of cosmic time. Cosmological perturbations may  gravitationally collapse to  form a PBH depending on the amplitude measured at the peak of the compaction function, defined as the mass excess compared to the background value within a given radius (see for instance Ref. \cite{Harada:2023ffo})

\be
C(\x)=2\frac{M(\x,t)-M_{\rm b}(\x,t)}{R(\x,t)},
\ee
where $M(\x,t)$ is the Misner-Sharp mass and   $M_{\rm b}(\x,t)$ its background value. The Misner-Sharp mass gives the mass within a sphere of
areal   radius 

\be
\label{R}
R(\x,t)=a(t)\tilde{r} e^{\zeta(\bf x)}
\ee
with spherical coordinate radius $\tilde{r}$, centered around
position $\x$  and evaluated at time $t$. The compaction directly measures
the overabundance of mass in a region and is therefore better suited than the curvature
perturbation for determining when an overdensity collapses into a PBH. 
Furthermore, the compaction has the advantage to be  time-independent
on superhorizon scales.  It  can be
written in terms of the density contrast as

\be
C(\x)=\frac{2\rho_{\rm b}}{R(\x,t)}\int {\rm d}^3\x  \,\delta(\x,t),
\ee
where $\rho_{\rm b}$ is the background energy density. On  superhorizon scales the density contrast  is related to the curvature
perturbation in real space by the nonlinear relation

\be
 \delta(\x,t)=-\frac{4}{9}\frac{1}{a^2H^2}e^{-2\zeta(\bf x)}\left(\nabla^2\zeta(\x)+\frac{1}{2}\left({\mathbf \nabla}\zeta(\x)\right)^2\right).
\ee
Assuming spherical symmetry and defining $\zeta'={\rm d}\zeta/{\rm d} r$,  the compaction function becomes

\begin{eqnarray}
\label{ccc}
C(r)&=&\frac{8\pi \rho_{\rm b}}{R(r,t)}\int_0^R{\rm d}\widetilde R\,\widetilde R^2(r,t)\delta(r,t)=C_\zeta(r)-\frac{3}{8}C_\zeta^2(r),\nonumber\\
C_\zeta(r)&=&-\frac{4}{3} r\zeta'(r).
\end{eqnarray}
Suppose now that there is peak in the curvature perturbation $\zeta(\x)$ with a given peak value $\zeta(0)$ and profile $\zeta(r)$ away from the center, which we arbitrarily can set at the origin of the coordinates. The corresponding compaction function will have  a maximum at the distance $r_m$ from the origin of the peak. Since

\be
\label{max}
C'(r_m)=C'_\zeta(r_m)\left[1-\frac{3}{4}C_\zeta(r_m)\right]=0,
\ee
the extremum of the compaction function $C(r)$ coincides with the extremum of $C_\zeta(r)$. Furthermore, since

\be
C''(r_m)=C''_\zeta(r_m)\left[1-\frac{3}{4}C_\zeta(r_m)\right],
\ee
the maximum  of the compaction function $C(r)$ coincides with the  maximum of $C_\zeta(r)$ as long as $C_\zeta(r_m)<4/3$ (the so-called type I case). We will focus therefore mainly  on this quantity. Notice that  sometimes  we will call $C_\zeta(r)$   ``linear" compaction function for simplicity, where the term linear stems from the fact that its expression is linear in the curvature perturbation $\zeta(r)$. However, $C_\zeta(r)$ is not necessarily Gaussian if the curvature perturbation $\zeta(r)$ is not.

The  maximum of the compaction function  is  fixed by the equation

\be
\label{sdd}
C'(r_m)=C'_\zeta(r_m)=0\,\,\,\,{\rm or}\,\,\,\, \zeta'(r_m)+ r_m\zeta''(r_m)=0.
\ee
Consider now a family of
compaction functions which have in common the same value of $r_m$, but a different curvature at the maximum   parametrized by \cite{Escriva:2019phb}

\be
q=-\frac{1}{4}\frac{r_m^2 C''(r_m)}{C(r_m)},
\ee
Numerically, it has been noticed that 
the critical threshold depends on the curvature at the peak of the compaction function \cite{Escriva:2019phb,musco,musconl}

\be\label{eq:Threshold}
C_{c}(q)=\frac{4}{15}e^{-1/q}\frac{q^{1-5/2q}}{\Gamma(5/2q)-\Gamma(5/2q,1/q)},
\ee
such that $
C_{c}(q\rightarrow 0)\simeq 2/5$ and  $
C_{c}(q\rightarrow \infty)\simeq 2/3$. We also notice  that 
\be
\label{notice}
q=-\frac{1}{4}\frac{r_m^2 C_\zeta''(r_m)\left[1-\frac{3}{4}C_\zeta(r_m)\right]}{C_\zeta(r_m)\left[1-\frac{3}{8}C_\zeta(r_m)\right]}\simeq -\frac{1}{4}\frac{r_m^2 C_\zeta''(r_m)}{C_\zeta(r_m)}\left[1-\frac{3}{8}C_\zeta(r_m)\right]\equiv q_\zeta\left[1-\frac{3}{8}C_\zeta(r_m)\right].
\ee

\subsection{The average profile}
\noindent
One question to pose is the following: which profile should one make use of to calculate the critical value for PBH abundance, given that it depends on the peak profile? The natural answer, routinely adopted in the literature,  would be the average profile of the compaction function with the constraint that  there is a peak of the curvature perturbation at the center of the coordinates with value $\zeta(0)$. This is the most obvious answer  as the average profile is the most frequent, statistically speaking. Supposing for the moment that $\zeta(r)$ is Gaussian, such an average profile would be

\be
\langle C_\zeta(r)\rangle_{\zeta(0)}=-\frac{4}{3}r
\langle \zeta'(r)\rangle_{\zeta(0)}=-\frac{4}{3}r\langle \zeta(r)\rangle_{\zeta(0)}'=-\frac{4}{3} r\frac{\xi'(r)}{\xi(0)}\zeta_0,
\ee
where 

\be
\xi(r)=\int\frac{{\rm d} k}{k}{\cal P}_\zeta(k) \frac{\sin k r}{k r}
\ee
is the two-point correlation of the curvature perturbation. In such a case the value of $r_m$ where the most likely compaction function has its maximum would then be fixed by the equation

\be
\xi'(r_m)+r_m\xi''(r_m)=0.
\ee
A standard choice is therefore to calculate the curvature of the peak of the compaction function as\footnote{This  is clearly not  correct as, for instance,  the average of the ratio of two stochastic variables is not the ratio of their averages.}

\be
q=-\frac{1}{4}\frac{r_m^2 \langle C''_\zeta(r_m)\rangle_{\zeta(0)}}{\langle C_\zeta(r_m)\rangle_{\zeta(0)}}\left[1-\frac{3}{8}\langle C_\zeta(r_m)\rangle_{\zeta(0)}\right].
\ee
The crucial point is  that, the smaller the value of the curvature, the smaller the value of the threshold. Since the PBHs abundance has an exponentially strong dependence on the threshold, one expects  that broad compaction functions should be very relevant in the determination of the abundance of PBHs even though they are more rare than the average profiles. This is what we discuss next.

\section{ The relevance of Broadness: the Gaussian case }
\noindent
In this section we  assume    $C_\zeta(r_m)$ and $C_\zeta''(r_m)$ to be  Gaussian (and correlated) variables. This will allow us to gain some analytical intuition.   We define

\be
 \sigma^2_0=\langle  C_\zeta^2(r_m)\rangle, \,\,\,\sigma^2_1= -\frac{1}{4}r_m^2\langle  C_\zeta''(r_m) C_\zeta(r_m)\rangle,\,\,\,{\rm and}\,\,\,\sigma^2_2=\frac{1}{16}r_m^4\langle {C_\zeta''(r_m)}^2\rangle.
\ee
Such correlations are easily computed knowing that the Fourier transform of the linear compaction function reads

\be
C_\zeta({\bf k},r)=\frac{4}{9}k^2 r^2 W(kr)\zeta({\bf k}),\,\,\,\,W(x)=3\frac{\sin x-x\cos x}{x^3},
\ee
where $W(x)$
is the Fourier transform of the Heaviside window function in real space.
We will  use the conservation of the probabilities 

 \begin{eqnarray}
     P\left[C(r_m),C''(r_m)\right]{\rm d}C(r_m){\rm d}C''(r_m)&=&
{\cal P}\left[C_\zeta(r_m),C_\zeta''(r_m)\right]{\rm d}C_\zeta(r_m){\rm d}C_\zeta''(r_m)\nonumber\\&=&\widetilde{{\cal P}}\left[C_\zeta(r_m),q_\zeta\right]{\rm d}C_\zeta(r_m){\rm d}q_\zeta
 \end{eqnarray}
where 
\begin{eqnarray}
{\cal P}\left[-\frac{1}{4}r_m^2  C_\zeta''(r_m),  C_\zeta(r_m)\right]&=&\frac{1}{2\pi\sqrt{{\rm det}\,\Sigma}}
\,{\rm exp}\left(-\vec V^T \Sigma^{-1}\vec V/2\right),\nonumber\\
\vec{V}^T&=&\left[-\frac{1}{4}r_m^2  C_\zeta''(r_m),  C_\zeta(r_m)\right],\nonumber\\
\Sigma&=&\left(\begin{array}{cc}
\sigma^2_2 & \sigma^2_1\\
\sigma^2_1 & \sigma^2_0\end{array}\right).
\end{eqnarray}
We find it convenient to define the  parameter

\be
\gamma=\frac{\sigma_1^2}{\sigma_2\sigma_0},
\ee
which  will play an important role in the following and indicates the broadness of a given power spectrum of the curvature perturbation. The 
closer $\gamma$ is to unity, the more spiky is the peak of the curvature perturbation.

\subsection{The average of the curvature}
\noindent
The average  curvature of the linear compaction function $C_\zeta$
 can   be computed by using the conditional probability to have a peak at $r_m$\footnote{In fact we use threshold statistics rather than peak statistics to elaborate our point. However, regions  well above the corresponding square root of the variance are very likely local maxima  \cite{Hoffman:1985pu}. }

\be
\langle q_\zeta\rangle=\int_0^\infty{\rm d}q_\zeta\,q_\zeta\,P[q|C_{\zeta}(r_m)>C_{\zeta,c}(q_\zeta)],
\ee
with 

\be
P[q|C_{\zeta}(r_m)>C_{\zeta,c}(q_\zeta)]=\frac{\widetilde{{\cal P}}[q_\zeta,C_{\zeta}(r_m)>C_{\zeta,c}(q_\zeta)]}{\widetilde{{\cal P}}[C_{\zeta}(r_m)>C_{\zeta,c}(q_\zeta)]}.
\ee
The conditional probability, in the limit of large thresholds, becomes

\begin{eqnarray}
    P[q_\zeta|C_{\zeta}(r_m)>C_{\zeta,c}(q_\zeta)]&\simeq &\frac{\left(1-\gamma^2\right)^{1/2}\sigma_2 C_{\zeta,c}(q_\zeta)}{\sqrt{2\pi}\sigma_0\left[(q_\zeta-\gamma\sigma_2/\sigma_0)^2+(1-\gamma^2)\sigma_2^2/\sigma_0^2\right]^{1/2}}\cdot
\nonumber\\
&\cdot& \exp\left[-\frac{(q-\gamma\sigma_2/\sigma_0)^2C^2_{\zeta,c}(q_\zeta)}{2(1-\gamma^2)\sigma_2^2}\right].
\end{eqnarray}
For a monochromatic power spectrum of the curvature perturbation, that is $\gamma\simeq 1$, we recognize the Dirac delta and the value of $q_\zeta$, which minimizes the exponent and maximizes the PBHs abundance, is the average value $\langle q_\zeta\rangle=\sigma_2/\sigma_0$. 

Departing from $\gamma\simeq 1$, and integrating numerically, one discovers departures from the value $\gamma\sigma_2/\sigma_0$ for the average of $q_\zeta$, but not dramatically, and one has

\be
\label{XY}
\langle q_\zeta\rangle\simeq \gamma\frac{\sigma_2}{\sigma_0}.
\ee
Hence for very broad spectrum, $\gamma \rightarrow 0$, one has $\langle q_\zeta\rangle \rightarrow 0$.
\subsection{The PBHs abundance}
\noindent
The PBHs abundance is  given by\footnote{We do not account for the extra factor counting the mass of the PBH with respect to the mass contained in  the horizon volume at re-entry as we give priority to getting analytical results. We will reintegrate it in the next section.}

\be
\beta=\int_{C_c(q)}^\infty{\rm d}C(r_m)\int_{-\infty}^0{\rm d}C''(r_m) P\left[C(r_m),C''(r_m)\right]=\int_{0}^\infty {\rm d}q_\zeta\int_{C_{\zeta,c}(q_\zeta)}^\infty{\rm d}C_\zeta(r_m) \widetilde{{\cal P}}\left[C_\zeta(r_m),q_\zeta\right].
\ee
Going back to the initial probability, it can be written as

\begin{eqnarray}
    {\cal P}\left[-\frac{1}{4}r_m^2  C_\zeta''(r_m),  C_\zeta(r_m)\right]&=&\frac{1}{2\pi}\frac{1}{\sigma_2\sigma_0\sqrt{1-\gamma^2}}{\rm exp}\left[-\frac{r_m^4 {C_\zeta''(r_m)}^2}{16\cdot 2\sigma_2^2}\right]\cdot\nonumber\\
    &\cdot &{\rm exp}\left[-\frac{1}{2(1-\gamma^2)}\left(\frac{C_\zeta(r_m)}{\sigma_0}+\gamma\frac{ r_m^2 C_\zeta''(r_m)}{4\sigma_2}\right)^2\right].
\end{eqnarray}
For a monochromatic, very peaked,  power spectrum of the curvature perturbation, where $\gamma\simeq 1$, the probability reduces to 

\be
\lim_{\gamma\rightarrow 1}{\cal P}\left[-\frac{1}{4}r_m^2  C_\zeta''(r_m),  C_\zeta(r_m)\right]=\frac{1}{\sqrt{2\pi}}\frac{1}{\sigma_2\sigma_0}{\rm exp}\left(-\frac{r_m^4 {C_\zeta''(r_m)}^2}{16\cdot 2\sigma_2^2}\right)\delta_D\left(\frac{C_\zeta(r_m)}{\sigma_0}+\frac{ r_m^2 C_\zeta''(r_m)}{4\sigma_2}\right),
\ee
which fixes 

\be
q_\zeta=\frac{\sigma_2}{\sigma_0}=\frac{\sigma_2\sigma_0}{\sigma_0^2}=\frac{\sigma^2_1}{\sigma^2_0}=\langle q_\zeta\rangle,
\ee
and

\begin{eqnarray}
\beta&=&\int_{0}^\infty{\rm d}q_\zeta\int_{C_{\zeta,c}(q_\zeta)}^{4/3}{\rm d}C_\zeta(r_m){\cal P}\left[-\frac{1}{4}r_m^2  C_\zeta''(r_m),  C_\zeta(r_m)\right]\nonumber\\
&=&\int_{C_{\zeta,c}(\langle q_\zeta\rangle)}^{4/3}{\rm d}C_\zeta(r_m)\frac{1}{\sqrt{2\pi}\sigma_0}{\rm exp}\left(-\frac{C^2_\zeta(r_m)}{2\sigma_0^2}\right)
=\frac{1}{2}{\rm Erfc}\left[\frac{C_{\zeta,c}(\langle q_\zeta\rangle)}{\sqrt{2}\sigma_0}\right],
\end{eqnarray}
where
\be
\label{c}
C_{\zeta,c}(q_\zeta)\simeq\frac{4}{3}\left(1-\sqrt{\frac{2-3C_{c}(q_\zeta) }{2}}\right).
\ee
Therefore for monochromatic spectra of the curvature perturbation the PBHs abundance is fixed by the value of the threshold corresponding to the average value of the curvature of the compaction function at its peak

\be
C^{\rm peaked}_{\zeta,c}=C_{\zeta,c}(\langle q_\zeta\rangle).
\ee
For a generic power spectrum, we change the variables from $(-r_m^2 C_\zeta''(r_m),C_\zeta(r_m))$ to $(q_\zeta,C_\zeta(r_m))$ and making use of the conservation of the probability we obtain

\begin{eqnarray}
\beta(q_\zeta)&=&\int_{C_{\zeta,c}(q_\zeta)}^{4/3}{\rm d}C_\zeta(r_m)|C_\zeta(r_m)|\,{\cal P}\left[q C_\zeta(r_m),  C_\zeta(r_m)\right]\nonumber\\
&\simeq& \frac{\sqrt{1-\gamma^2}\sigma_2}{2\pi\sigma_0\left[(q_\zeta-\gamma \sigma_2/\sigma_0)^2+(1-\gamma^2)\sigma^2_2/\sigma^2_0\right]}\cdot\nonumber\\
&\cdot&\exp\left[-\frac{\left[(q_\zeta-\gamma \sigma_2/\sigma_0)^2+(1-\gamma^2)\sigma^2_2/\sigma^2_0\right]C^2_{\zeta,c}(q_\zeta)}{2(1-\gamma^2)\sigma^2_2}\right].\nonumber\\
&&
\end{eqnarray}
We see that the square of the critical threshold is replaced by an effective squared critical threshold
\vskip 0.3cm

\begin{figure}[htb]
\begin{minipage}[t]{0.45\linewidth}
\includegraphics[width=\linewidth]{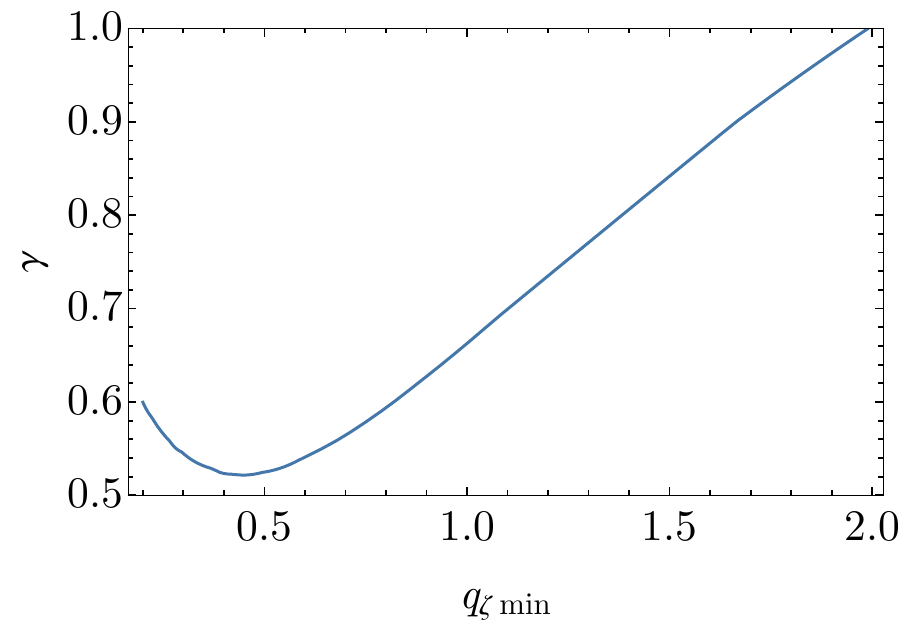}
\caption{Plot of $q_{\zeta \; \rm{\min}}$ as a function of $\gamma$ for $\sigma_2/\sigma_0=2$.}
\label{fig:gamma_qmax}
\end{minipage}
\hfill
\begin{minipage}[t]{0.45\linewidth}
\includegraphics[width=\linewidth]{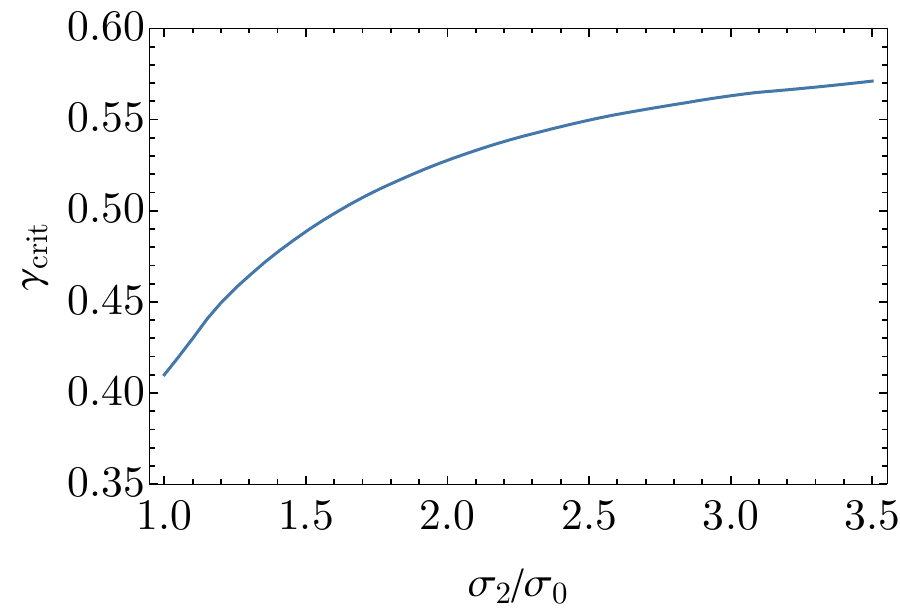}
\caption{Critical value of $\gamma$ as a function of $\sigma_2/\sigma_0$}
\label{fig:gamma_crit}
\end{minipage}%
\end{figure}

\begin{figure}[hbt]
\centering
  \includegraphics[width=14cm]{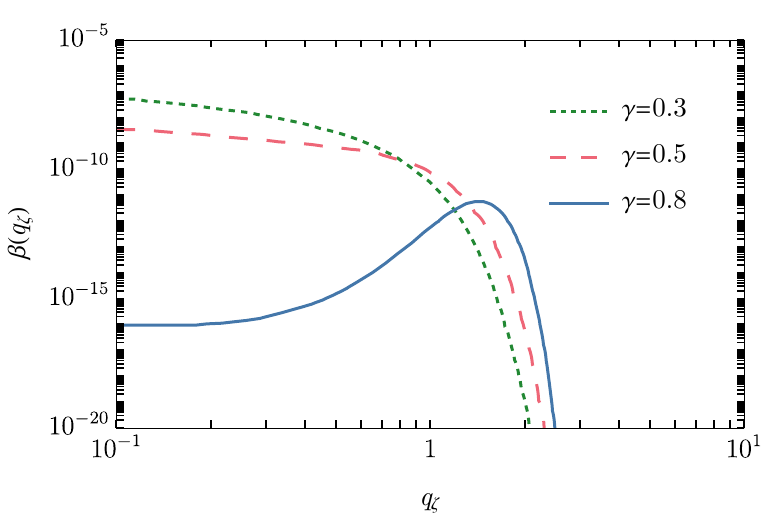}
  \caption{The PBHs formation probability as a function of $q_\zeta$ for $\sigma_0=\sigma_2/2=0.05$ and three different values of $\gamma=(0.3,0.5,0.8)$ for which $\langle q_\zeta\rangle =(0.6,1,1.6)$ for the Gaussian case.}
  \label{fig:three_gammas}
\end{figure}

\begin{mdframed}
\be
C^2_{\zeta,c}(q_\zeta)\Big|_{\rm eff}
=\left[(q_\zeta-\gamma \sigma_2/\sigma_0)^2+(1-\gamma^2)\sigma_2^2/\sigma_0^2\right] C^2_{\zeta,c}(q_\zeta).
\ee
\vskip 0.3cm
\end{mdframed}
Its minimum is determined by the equation 
\be
\label{min}
(q_\zeta-\gamma\sigma_2/\sigma_0)C_{\zeta,c}(q_\zeta)+\left[(q_\zeta-\gamma\sigma_2/\sigma_0)^2+(1-\gamma^2)\sigma_2^2/\sigma_0^2\right]\frac{{\rm d}C_{\zeta,c}(q_\zeta)}{{\rm d}q_\zeta}=0.
\ee
For peaked profiles where $\gamma\simeq 1$ we have

\be
(q_\zeta-\sigma_2/\sigma_0)\left[C_{\zeta,c}(q_\zeta)+(q_\zeta-\sigma_2/\sigma_0)\frac{{\rm d}C_{\zeta,c}(q_\zeta)}{{\rm d}q_\zeta}\right]=0
\ee
and the mimimum lies at the value of the average $q_\zeta=\langle q_\zeta\rangle=\sigma_2/\sigma_0$. For broad spectra $\gamma\ll 1$, the effective threshold is minimized for $q_\zeta\simeq 0$ as it reduces to 

\be
\label{eq:large_q0}
C^2_{\zeta,c}(q_\zeta)\Big|_{\rm eff}
=\left[q_\zeta^2+\sigma_2^2/\sigma_0^2\right]C^2_{\zeta,c}(q_\zeta).
\ee
and the threshold $C_{\zeta,c}(q_\zeta)$ is also minimized for small $q_\zeta$. There is in general a critical value of $q_\zeta$ for which the abundance is always dominated by the broad spectra. We can see this behaviour by plotting the curve $(q_{\zeta\;\rm{min}},\gamma)$ obtain from the Eq.\,(\ref{min}), as shown in Fig.\,\ref{fig:gamma_qmax}. As we start decreasing from $\gamma=1$ where the minimum is in $\sigma_2/\sigma_0$, also the value of $q_{\zeta,\;\rm{min}}$ decreases, up until a critical value $\gamma_{\rm crit}$.
For values of $\gamma$ below this point the function $C_{\zeta,c}(q_{\zeta})|_{\rm{eff}}$ does not have a minimum, but is monotonically increasing with $q_{\zeta}$, hence the minimum lies at the boundary of the interval, i.e. $q_{\zeta}=0$. The transition is therefore very sharp after  the critical value.\\
It is also possible to evaluate the position of this minimum for different values of the parameter $\sigma_2/\sigma_0$, as shown in Fig. \ref{fig:gamma_crit}. We can understand the behaviour because having larger values of this parameter the transition happens for larger values of $\gamma$, being easier to enter in the regime of Eq. (\ref{eq:large_q0}), where $\sigma_2/\sigma_0$ dominates.
 To show this explicitly, in Fig. \ref{fig:three_gammas} we plot the formation probability for  three different values of $\gamma$. It demonstrates that the abundance is dominated by the broadest profiles when the curvature perturbation is not very spiky and not by the average value of $q_\zeta$.
The corresponding critical value needed to be used is therefore

\be
\label{cc}
C_{\zeta,c}(q_\zeta\simeq 0)\simeq\frac{4}{3}\left(1-\sqrt{\frac{2-3\cdot 2/5 }{2}}\right)\simeq 0.49.
\ee
\noindent

\section{ The relevance of Broadness: the non-Gaussian case }
\noindent
As a matter of fact, the curvature perturbation generated in models producing large overdensities is typically non-Gaussian. Non-Gaussianity among the modes interested in the growth of the curvature perturbation is generated either by their self-interaction during the ultra slow-roll phase \cite{cai} or after Hubble radius exit when the curvature perturbation is sourced by a curvaton-like field \cite{curvaton, curvaton_2,curvaton_3}.

We proceed therefore by assuming that  the initial curvature perturbation is non-Gaussian, but a function of a Gaussian component 

\be
\zeta(r)=F[\zeta_{\rm g}(r)].
\ee
In such a case the compaction function is still given by Eq. (\ref{ccc}), where 

\be
C_\zeta(r)= F_1(\zeta_{\rm g}) C_{\rm g}(r),\,\,\,\,
C_{\rm g}(r)=-\frac{4}{3}r\zeta'_{\rm g}(r),
\ee
and we have indicated the derivatives of $F$ with respect to $\zeta_\g$ by $F_n =\dd F(\zeta_\g)/ \dd \zeta_\g$.
 The maximum of the compaction function can be found solving the equation
\begin{equation}
    C_\zeta'(r_m) = F_1(\zeta_\g) C'_{\g}(r_m) + C_{\g}(r_m) \zeta_\g'(r_m)  F_2(\zeta_\g) = 0,
\end{equation}
as long as $C_\zeta(r_m)<4/3$.
The next  step is to define  the  following  Gaussian and correlated variables  
\begin{equation}
    x_0 = \zeta_\g, \quad x_1 = r\zeta'_\g, \quad x_2 = r^2\zeta''_\g, \quad x_3 =r^3\zeta'''_\g,
\end{equation}
for which the condition of the maximum becomes
\begin{equation}
    x_2 = -x_1\left(1 + x_1 \frac{F_2(x_0)}{F_1(x_0)}\right).
\end{equation}
One can construct the corresponding   probability distribution as 
\begin{equation}
    P(x_0,x_1,x_2,x_3) = \frac{1}{(2\pi)^2\sqrt{{\rm det}\,\Sigma}}
\,{\rm exp}\left(-\vec V^T \Sigma^{-1}\vec V/2\right), 
\end{equation}
where
\begin{equation*}
    \vec{V}^T= \left[x_0,x_1,x_2,x_3\right],
\end{equation*}
and
\begin{eqnarray}
\Sigma&=&\left(\begin{array}{cccc}
\sigma^2_0 & \gamma_{01}\sigma_1\sigma_0 &\gamma_{02}\sigma_2\sigma_0 &\gamma_{03}\sigma_3\sigma_0\\
\gamma_{01}\sigma_1\sigma_0 & \sigma_1^2 &\gamma_{12}\sigma_2\sigma_1 &\gamma_{13}\sigma_1\sigma_3\\
 \gamma_{02}\sigma_2\sigma_0 & \gamma_{12}\sigma_2\sigma_1 &\sigma^2_2&\gamma_{23}\sigma_2\sigma_3\\
\gamma_{03}\sigma_3\sigma_0& \gamma_{13}\sigma_1\sigma_3 &\gamma_{23}\sigma_2\sigma_3 &\sigma_3^2\end{array}\right)
\end{eqnarray}
is constructed from the different correlators with\footnote{Notice that here for clarity the index for the various $\sigma_i$ is related to the number of derivatives of $\zeta_\g$, differently from the definition in the previous section.}
\begin{equation}
    \sigma_i^2 = \langle x_i^2\rangle, \quad \gamma_{ij}=\frac{\langle x_i x_j\rangle}{\langle x_i^2 \rangle^{1/2}\;\langle x_j^2 \rangle^{1/2}}.
\end{equation}
Next, we need to  convert all the relevant variables in terms  of the Gaussian ones   $x_i$ ($i=0,\cdots,3)$. 
First we have
\begin{equation}
    C_{\g} = -\frac{4}{3} \, x_1,
\end{equation}
and the  derivatives of $C_\zeta$ can be written in terms of $x_1$ and $x_2$ as
\begin{eqnarray}
    C_{\zeta} &=& -\frac{4}{3}\; x_1 F_1(x_0), \nonumber\\
    rC_{\zeta}' &=& -\frac{4}{3} \left(F_1(x_0) (x_1 + x_2) + x_1^2 F_2(x_0) \right), \nonumber\\
    r^2C_{\zeta}'' &=& -\frac{4}{3} \left[ F_1(x_0) (2x_2+ x_3) + 2 x_1^2 F_2(x_0) + 3 x_1x_2 F_2(x_0) + x_1^3 F_3(x_0)\right].
\end{eqnarray}
The  PBHs  abundance for a given value of the curvature $q$ will read 

\begin{equation}
    \beta(q) = \int_{D}\mathcal{K}\left(C - C_c(q)\right)^{\gamma}\; p(x_0, C_{\zeta}, x_2, q)\;\delta\left( F_1(x_0) (x_1
    + x_2) + x_1^2 F_2(x_0) \right),
\end{equation}
where the domain of integration is 

\begin{equation}
    D = \left\{x_2\in \mathbb{R}, C_{\zeta,c}(q)<C_\zeta<\frac{4}{3}\right\},
\end{equation}
with 
\be
\label{c1}
C_{\zeta,c}(q)\simeq\frac{4}{3}\left(1-\sqrt{\frac{2-3C_{c}(q) }{2}}\right).
\ee
We have reintroduced the  scaling-law factor for critical collapse $\mathcal{K}(C-C_c(q))^\gamma$  which  accounts for the mass of the PBHs at formation written in units of the horizon mass at the time of horizon re-entry, with  $\mathcal{K}\simeq  3.3$  for a log-normal power spectrum and $\gamma\simeq 0.36$ \cite{Choptuik_gamma, crit_1,crit_2,Ianniccari:2024ltb} (see also Ref. \cite{Sasaki_review}). By using the conservation of probabilities we can finally write 
 \be
p\left[\zeta_\g, C_\zeta,x_2,q\right]=P\left[x_0,x_1,x_2,x_3\right]|{\rm Det}\,  J|,
\ee
where 
\begin{equation}
    {\rm Det}\, J  = \frac{3}{4}\left(\frac{4x_1 +2F_1(x_0)x_1^2}{1+F_1(x_0)x_1}\right),
\end{equation}
and at the maximum
\begin{equation}
    x_3= \frac{-4q(1+\frac{1}{2}x_1F_1(x_0))x_1}{1+x_1F_1(x_0)}-2x_2-2x_1^2\frac{F_2(x_0)}{F_1(x_0)}-3x_1x_2\frac{F_2(x_0)}{F_1(x_0)}-x_1^3\frac{F_3(x_0)}{F_1(x_0)}.
\end{equation}
We rewrite the Gaussian probability in the following form 
\be
P(x_0, x_1, x_2, x_3) = \frac{1}{4\pi^2 \sqrt{\det{\Sigma}}} \exp{\left(-\frac{(\sigma_0 \sigma_1 \sigma_2 \sigma_3)^2}{2\det{\Sigma}} \sum_{i,j=0}^3 \frac{\kappa_{ij} x_i x_j}{\sigma_i\sigma_j} \right)},
\ee
where the $\kappa_{ij}$'s will depend on all the $\gamma_{lm}$, and they can be computed by performing the inverse of the matrix $\Sigma$, matching with the definition. 
Performing the change of variables we get 
\begin{eqnarray}
    p(\zeta_\g, C_{\zeta}, x_2, q) &=&  \left|\frac{9 C_\g \left(3 C_\g F_1-8\right)}{8 \
\left(3 C_\g F_1-4\right)} \right| \frac{1}{4\pi^2 \sqrt{\det{\Sigma}}}\cdot\nonumber\\
&\cdot& \exp{\left(-\frac{(\sigma_0 \sigma_1 \sigma_2 \sigma_3)^2}{2\det{\Sigma}} \frac{1}{4096 \
F_1^4}\left[ A(\zeta_\g, C_\g) q^2 + B(\zeta_\g, C_\g) q + C(\zeta_\g, C_\g) \right] \right)},\nonumber\\
&&
\end{eqnarray}
where we have defined the following functions of $\zeta_\g$ and $C_\g$
\be
    A(\zeta_\g, C_\g)=\frac{9216 \kappa_{33} C_\g^2 F_1^4 
\left(8-3 C_\g F_1\right){}^2}{\sigma _3^2 
\left(4-3 C_\g F_1\right){}^2},
\ee
\begin{eqnarray}
B(\zeta_\g, C_\g)&=&\frac{192 C_\g F_1^2 \left(3 C_\g 
F_1-8\right)}{\sigma _0 \sigma _1 \sigma _2 
\sigma _3^2 \left(3 C_\g F_1-4\right)} \left[3 \kappa _{33} \sigma _0 
\sigma_1 \sigma _2 C_\g \left\{9 C_\g F_1 
\left(C_\g F_3+4 F_2\right)-27 C_\g^2 F_2^2-32 
F_1^2\right\}\right. \nonumber\\
&&\left. +4 \sigma _3 
F_1 \left\{3 \kappa _{23} \sigma_0 \sigma_1 C_\g \left(4 F_1-3 C_\g 
F_2\right)+4 \sigma_2 F_1 \left(4 \kappa _{30} \sigma_1 \zeta_\g-3 \kappa _{13} 
\sigma_0 C_\g\right)\right\}\right],
\end{eqnarray}
\begin{eqnarray}
C(\zeta_\g, C_\g)&=&-\frac{\kappa_{33}}{\sigma_3^2}\left\{ 4374  C_\g^6 F_1 
F_2^2 F_3 - 5832  C_\g^5 
F_1^2 F_2 
F_3+ 17496  C_\g^5 
F_1 F_2^3- 27216  C_\g^4 
F_1^2 F_2^2+\right.\nonumber\\
&&\left.20736  C_\g^3 
F_1^3 F_2 - 729  
C_\g^6 F_1^2 F_3^2+ 5184  C_\g^4 
F_1^3 F_3-9216  C_\g^2 
F_1^4-6561  C_\g^6 F_2^4 \right\}+\nonumber\\
&& -\frac{1}{\sigma _0 \sigma _1 
\sigma _2 \sigma _3}24 C_\g \
F_1 \left(-9 C_\g F_1 \left(C_\g F_3+4 
F_2\right)+27 C_\g^2 F_2{}^2+32 F_1^2\right)\nonumber\\
&&\left(3 
\kappa _{23} \sigma_0 \sigma _1 C_\g \left(4 F_1-3 C_\g F_2\right)+4 \sigma _2 
F_1 \left(4 \kappa _{30} \sigma _1 \zeta_\g-3 \kappa _{13} \sigma _0 C_\g\right)\right)+\nonumber\\
&&+16 F_1^2 
\left(\frac{24 C_\g F_1 \left(3 C_\g 
F_2-4 F_1\right) 
\left(3 \kappa _{12} \sigma _0 C_\g-4 \kappa _{20} \sigma _1 \zeta_\g\right)}{\sigma _0 \sigma _1 \sigma _2}+\right.\nonumber\\
&&\left. +\frac{9 \kappa _{22} C_\g^2 \
\left(4 F_1-3 C_\g F_2\right){}^2}{\sigma _2^2}+\frac{16 F_1^2 \left(-24 \kappa _{10} \sigma _1 \sigma _0 C_\g \zeta_\g+9 \kappa _{11} \sigma_0^2 C_\g^2+16 \kappa_0 \sigma _1^2 \zeta_\g^2\right)}{\sigma _0^2 \sigma _1^2}\right),\nonumber\\
&&
\end{eqnarray}
and  each  function $F_n$ is intended to be $F_n(\zeta_\g)$.

\subsection{An illustrative example}
\begin{figure}
    \centering
    \includegraphics[width=14cm]{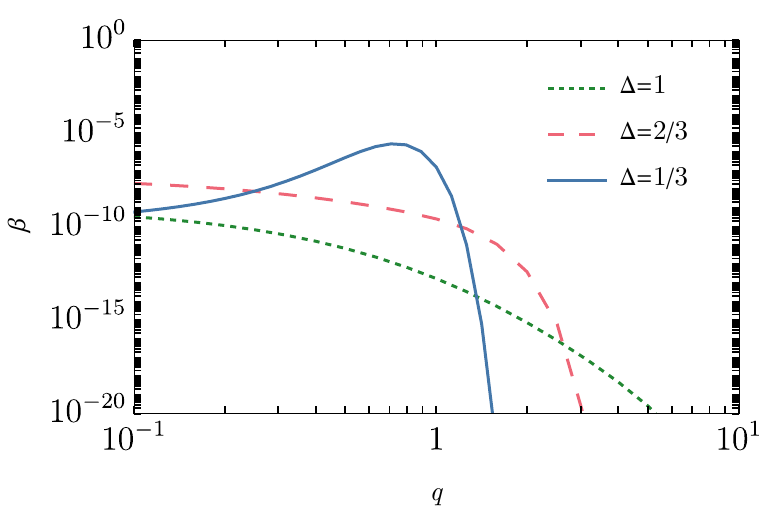}
    \caption{Mass fraction $\beta$ for the non-Gaussian scenario computed with several values of $\Delta$, where we fix $\mu_* = 5/2$ and the amplitude of the power spectrum $A=10^{-2}$.} 
    \label{fig:nonlin_gamma_}
\end{figure}

\noindent
We consider the following illustrative example which typically arises in models in which the curvature perturbation is generated during a period of ultra-slow-roll \cite{cai,b1,b2,Tomberg:2023kli}\footnote{For $\zeta_\g> \mu_\star$, Eq. (\ref{n}) does not capture the possibility of PBHs formed by bubbles of trapped vacuum which requires a separate discussion \cite{Escriva:2023uko,Uehara:2024yyp}.}

\be
\label{n}
\zeta({\bf x})=-\mu_\star\ln\left(1-\frac{ \zeta_\g({\bf x})}{\mu_\star}\right),
\ee
with $\mu_\star$ a model-dependent parameter depending upon the transition between the ultra-slow-roll phase and the subsequent slow-roll phase. To focus only on the impact of primordial non gaussianity, in this analysis we take $\mu_\star$ as a free parameter. We take the power spectrum of the Gaussian component to be a log-normal power spectrum

\be\label{eq:PS}
{\cal P}_\g(k)=\frac{A}{\sqrt{2\pi}\Delta}{\rm exp}\left[-\ln^2(k/k_\star)/2\Delta^2\right].
\ee
Our results are summarized in Fig.\,\ref{fig:nonlin_gamma_} where, changing the shape of the power spectrum, we computed  $k_\star r_m$ following ref\,\cite{musconl}\footnote{Here we stress that the value of $r_m$ can slighlty change in presence of large primordial non-guassianities\,\cite{Kehagias:2019eil}.}.
The  broadness of the power spectrum is controlled by the parameter  $\Delta$. We observe that by increasing the value of $\Delta$, enlarging the power spectra, again the PBHs formation probability is dominated by the broadest profiles. We have checked that for very peaked power spectrum, as in the case for $\Delta=1/3$, the abundance is peaked again around the average of $q$.

\section{Comparison with literature and impact on the physics of PBHs and Pulsar Timing Arrays}
\noindent
In this section we compare the calculation presented above, accounting for the curvature of the compaction function at its peak, with the prescription based on threshold statistics on the compaction function, reported in Refs. \cite{F,Gow}, where the only explicit dependence on $q$ is encoded in $C_c(q)$.
There, the  formation probability
is computed by integrating the joint probability distribution function $P_{\mathrm{g}}$ 
\be\label{eq:beta}
    \beta = 
    \int_{D}\mathcal{K}\left(C - C_c(q)\right)^{\gamma}
    \textrm{P}_{\rm g}(C_{\rm g},\zeta_{\rm g})\td C_{\rm g} \td\zeta_{\rm g}\,,
\ee
where the domain of integration is given by $D =
\left\{
    C(C_{\rm g},\zeta_{\rm g}) > C_{\rm c}(q), ~C_\zeta(C_{\rm g},\zeta_{\rm g}) < 4/3
\right\}
$. The Gaussian components are distributed as
\be
    P_{\rm g}\left(C_{\rm g}, \zeta_{\rm g}\right)
    = \frac{1}{2 \pi \sigma_{\rm a} \sigma_{\rm c} \sqrt{1-\gamma_*^2}} \exp \left[-\frac{1}{2\left(1-\gamma_*^2\right)}\left(\frac{C_{\mathrm{g}}}{\sigma_{\rm a}}-\frac{\gamma_* \zeta_{\mathrm{g}}}{\sigma_{\rm c}}\right)^2 \!-\! \frac{\zeta_{\mathrm{g}}^2}{2 \sigma_{\rm c}^2}\right],
\ee
with correlators 
\begin{subequations}
\begin{align}
  \langle C^2_{\rm g}\rangle & =  \sigma_{\rm a}^2=\frac{16}{81} \int_0^{\infty} \frac{\td k}{k}\left(k r_m\right)^4 W^2\left(k, r_m\right)T^2\left(k, r_m\right)P_\zeta, \\
    \langle C_{\rm g}\zeta_{\rm g}\rangle & =\sigma_{\rm b}^2=\frac{4}{9} \! \int_0^{\infty} \!\! \frac{\td k}{k} \!\left(k r_m\right)^2 \!W\!\!\left(k, r_m\right) \!W_s\!\left(k, r_m\right) T^2\left(k, r_m\right)\!P_\zeta, \\
     \langle \zeta^2_{\rm g}\rangle & =\sigma_{\rm c}^2=\int_0^{\infty} \frac{\td k}{k} W_s^2\left(k, r_m\right)T^2\left(k, r_m\right) P_\zeta,
\end{align}
\end{subequations}
and $\gamma_*=\sigma_{\rm b}^2/\sigma_{\rm a}\sigma_{\rm c}.$ We have defined $W\left(k, r_m\right)$ and $W_s\left(k, r_m\right)$ 
as the top-hat window function and the spherical-shell window function\,\cite{Young}. To compare this prescription with the one presented in this paper, we consider two cases: $\beta_0$, in which we do not adopt any transfer function ($T=1$) since everything is determined on superhorizon scales, and $\beta_T$, in which we consider the radiation transfer function assuming a perfect radiation fluid, as adopted in Ref. \cite{F}.

\begin{figure}
    \centering
    \includegraphics[width=14cm]{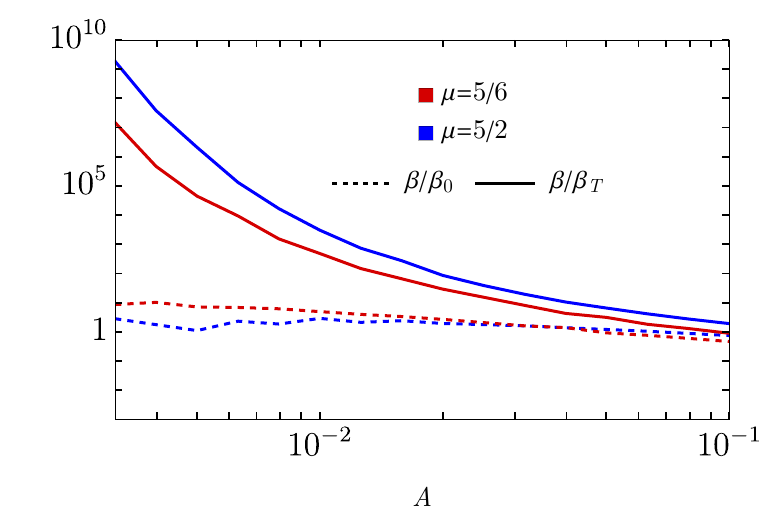}
    \caption{Ratio between mass fraction $\beta$ for the non-Gaussian case between the prescriptions presented in this paper and the prescription presented in Ref. \cite{F}. We fix the shape parameter $q=0.5$ (as a consequence also the threshold using Eq. (\ref{eq:Threshold})) and the shape of power spectrum $\Delta=0.5$ while we vary the amplitude.} 
    \label{fig:compa}
\end{figure}

In Fig. \ref{fig:compa}, we show a comparison between the two prescriptions using the typical non-Gaussian relation in the ultra-slow-roll scenario (see Eq. (\ref{n})) with a log-normal power spectrum (see Eq. (\ref{eq:PS})) with several benchmark values for $\mu_\star$. We fix $\Delta=1$ in the plots, but we have found analogous results also varying this parameter.
As we can understand from Fig. \ref{fig:compa}, evaluating the quantities on superhorizon scales, i.e. the ratio $\beta/\beta_0$, there is a marginal discrepancy between the two prescription. This discrepancy arises because, unlike the prescription used in the literature, where an average profile is employed, the effective threshold is slightly different than  the averaged case,  as evident from Eq.  (\ref{eq:large_q0}). Nevertheless an equivalent amount of PBHs requires a marginal change in the amplitude of the curvature perturbation power spectrum.

The situation is different when we include the radiation transfer function, i.e. the ratio $\beta/\beta_T$. The presence of the transfer function decreases the values of the variances and, as a consequence, it reduces the amount of PBHs.

This has important implications for the phenomenology related to PBHs respect to the case discussed in this paper.
Indeed in the standard formation scenario   PBH formation occurs as large curvature perturbations re-enter the Hubble horizon after inflation and eventually collapse under the effect of gravity. When such scalar perturbations cross the horizon they produce tensor modes as a second-order effect, which appear to us today as a signal of stochastic gravitational waves background (SGWB) (for a recent review, see Ref.~\cite{Domenech:2021ztg}). Recently in Ref.\cite{IovinoPTA}, where the old prescription was used, it was shown that large negative non-Gaussianity are necessary in order to achieve high enough amplitude, without overproducing PBHs, in order to relax the tension between the PTA recent dataset (the most constrained dataset is the one released by NANOGrav \cite{NANOGrav}) and the PBH explanation. 

We demonstrate that even when correctly accounting for the impact of the curvature of the compaction function and calculating  all the relevant quantities on  superhorizon scales, thereby avoiding all concerns regarding non-linearities in the radiation transfer function and the determination of the true physical horizon, the tension  between the PTA dataset and the PBH hypothesis is even worse than what claimed in Ref.\,\cite{IovinoPTA}.

We conclude this section, making a general comparison with another statistical approach for computing the PBH abundance. 
When the abundance is exponentially sensitive on the threshold, as well as in the case of peak theory\,\cite{Bardeen:1985tr,GreenPeak,Yoo:2018kvb,Yoo:2019pma,Yoo:2020dkz,Young:2020xmk,Kitajima:2021fpq,Taoso:2021uvl,Riccardi:2021rlf,Young} (see for example section 3 of ref.\,\cite{Yoo:2019pma} for a general expression of the PBH fraction in the context of peak theory or simply Eq. (32) and Eq. (33) of ref.\,\cite{Young}) we expect that our results can be generally extended to these other approaches. However there are still discrepancies between these two approaches, which are already present at Gaussian level. Indeed the approach based on peak theory requires slightly smaller values of the amplitude in order to get the same abundance of PBHs\,\cite{Young1,DeLuca1}, thus making the claim on the tension with the PTA dataset, even stronger. We leave a deeper analysis for the discrepancies between threshold statistics and peak theory in presence of primordial non-gaussianities for a future work.
\section{ Conclusions and some further final considerations}
\noindent
In this paper we have shown that the abundance of PBHs is dominated by the broadest profiles of the compaction function, even though they are not the typical ones, unless the power spectrum of the curvature perturbation is very peaked. The corresponding threshold is therefore always 2/5.
We have also discussed how this result makes the tension between overproducing PBHs and fitting the recent PTA data on gravitational waves even worse than recent analysis.

On more general grounds, given the dependence of the critical threshold on the profile of the compaction function, the natural question is if it possible to construct an observable whose critical threshold does not depend at all on the profiles of the peaks. 
In Ref. \cite{Escriva:2019phb} it has been proven numerically  that the volume average of the compaction function, calculated in a volume of sphere of radius $R_m$

    \be
\overline{C}(R_m)=\frac{3}{R_m^3}\int_0^{R_m}{\rm d} x\, x^2 \, C(x)
    \ee
     has a critical threshold equal to 2/5 independently from the profile. In the case of a broad compaction function, whose critical threshold is 2/5, and since
     $\overline{C}(R_m)\simeq C(R_m)$, it is trivial that the volume average has the same critical value 2/5. The case of a very spiky compaction function corresponds to a flat universe with in it a sphere of radius $R_m$ and constant curvature $K(R)=C(R)/R^2$, that is $C(R)$ scales like $R^2$. One then obtains 
     
     \be
\overline{C}(R_m)=3\frac{C(R_m)}{R_m^5}\int_0^{R_m}{\rm d} x\, x^4=\frac{3}{5}
    C(R_m)=\frac{3}{5}\cdot\frac{2}{3}=\frac{2}{5},
    \ee
     where it is used that for very spiky compaction functions the critical value is 2/3.

Assuming a universal threshold, one can then write  the probability that the  volume average compaction function is larger than 2/5  even for the non-Gaussian case as (we use here threshold statistics to make the point, one could similarly use peak theory)\footnote{For non-Gaussian perturbations the universal threshold remains 2/5 \cite{lo} for the realistic cases in which the non-Gaussian parameter is positive \cite{fr}.
Notice that one can construct easily another observable whose threshold
     is independent from the profile. 
     Indeed, as we mentioned already, the compaction function is related to the local curvature of the universe by the relation $C(R)=K(R)R^2$. Given a curvature perturbation $\zeta(r)$, a compaction function $C(R)$ with maximum in $R_m$ and the corresponding curvature $K(R)$,  one  consider a new perturbation with curvature
$$
\overline{K}=\Theta_H(R_m-R) \int_0^{R_m}{\rm d} x\, x^2 \, K(x),
     $$  
     that is a spherical local closed universe with curvature $\overline{K}$ with radius $R_m$ surrounded by a flat universe. This corresponds to a new  infinitely peaked compaction function equal to $\Theta_H(R_m-R)R^2\overline{K}$ whose threshold will be  always 2/3 \cite{HKY,Kehagias:2024kgk}, independently from the profile of the initial compaction function.}

\begin{eqnarray}
    \label{ppp}
    P\left[\overline{C}(R_m)>2/5\right]&=&\Big<\Theta_H\left[\overline{C}(R_m)-2/5\right]\Big>\nonumber\\
    &=&\frac{1}{2\pi}
\int\left[D C(r)\right]P\left[ C(r)\right]
\int_{2/5}^\infty{\rm d}\alpha
\int_{-\infty}^\infty {\rm d}\phi\,e^{i\phi(\overline{C}(R_m)-\alpha)}
\end{eqnarray}
which can be written as
\be
P\left[\overline{C}(R_m)>2/5\right]=
\int_{2/5}^\infty{\rm d}\alpha
\int_{-\infty}^\infty {\rm d}\phi\,e^{-i\phi\alpha}\cdot Z[J],
\ee
with
\begin{eqnarray}
 Z[J]=\int\left[D C({\bf x})\right]P\left[ C({\bf x})\right]e^{i\int {\rm d}^3 x\, J({\bf x}) C({\bf x})},\quad J({\bf x})=\,V^{-1}_{R_m}\,\phi\,\Theta_H (\orm-r)\, ,
\end{eqnarray}
and the measure $\left[D C(r)\right]$  is such that 
\begin{eqnarray}
\int\left[D C({\bf x})\right]P\left[ C({\bf x})\right]=\int\left[D C(r)\right]P\left[ C(r)\right]=1. 
\end{eqnarray}
The correlators are determined by the  expansion of the  partition function $Z[J]$  in terms of the source $J$,  while the corresponding expansion of $W[J]=\ln Z[J]$ generates the connected correlation functions. We will denote the latter as 
\begin{eqnarray}
    \xi^{(n)}({\bf x}_1,\cdots,{\bf x}_n)=
    \frac{\delta}{\delta J({\bf y}_1)}\cdots\frac{\delta}{\delta J({\bf y}_n)}\ln Z[J],
\end{eqnarray}
and the connected cumulants of the volume average linear compaction function as 
\begin{eqnarray}
\langle\overline{C}^n(R_m)\rangle&=&\frac{1}{V^{n}_{R_m}}
\int {\rm d}^3 x_1\cdots {\rm d} ^3 x_n\prod_{i=1}^n \xi^{(n)}({\bf x}_1,\cdots,{\bf x}_n)
\, \Theta_H(R_m-x_i)\nonumber\\
&=&\prod_{i=1}^n\int\frac{{\rm d}^3 k_i}{(2\pi)^3}P_N({\bf k}_1,\cdots,{\bf k}_n)W(k_1R_m)\cdots W(k_n R_m)\,\delta_D^{(n)}({\bf k}_1+\cdots+{\bf k}_n),\nonumber\\
\langle C_\zeta({\bf k}_1),\cdots,C_\zeta({\bf k}_n)\rangle &=& P_N({\bf k}_1,\cdots,{\bf k}_n) \delta_D^{(n)}({\bf k}_1+\cdots+{\bf k}_n).
\end{eqnarray}
Then, we may write 
\begin{eqnarray}
    \ln Z[J]&=& \sum_{n=2}^\infty\frac{(-1)^n}{n!}\int {\rm d}^3 {\bf y}_1\cdots \int {\rm d}^3 {\bf y}_n \,
 J_{i_1}({\bf y}_1)\cdots J_{i_n}({\bf y}_n)
 \xi^{(n)}({\bf x}_{i_1},\cdots,{\bf x}_{i_n})\nonumber\\
 &=& \sum_{n=2}^\infty\frac{(-1)^n}{n!}
\phi^n \langle\overline{C}^n\rangle .
\end{eqnarray}
	Using the above expression for the connected partition function, we find that the one-point statistics of Eq. (\ref{ppp}) can be written as 
	\begin{equation}
\begin{aligned}
\label{11}
P\left[\overline{C}(R_m)>2/5\right]
 =&
(2\pi)^{-1/2}\int_{2/5}^\infty {\rm d} a
\,
\exp\left\{\sum_{n=3}^\infty \frac{(-1)^n}{n!} 
\langle\overline{C}^n\rangle
\frac{\partial^n}{\partial a^n}\right\}
\exp{\left(-\frac{a^2}{2\sigma_{\overline{C}}^2}\right)}\\
=&
(2\pi)^{-1/2}\int_{2/5}^\infty {\rm d} a
\,\left(1-\frac{1}{3!}\langle\overline{C}_\zeta^3\rangle\frac{{\rm d}^3}{{\rm d} a^3}+
\frac{1}{4!}\langle\overline{C}^4\rangle\frac{{\rm d}^4}{{\rm d}a^4}+\cdots\right)\exp{\left(-\frac{a^2}{2\sigma_{\overline{C}}^2}\right)}\\
=& h_0(2/5)+\frac{1}{\sqrt{2\pi}}\sum_{n\geq 3}\frac{1}{2^{\frac{n}{2}}n!}\frac{c_n}{\sigma_{\overline{C}}^{n-1}}e^{-4/50\sigma_{\overline{C}}^2}H_{n-1}\left(\frac{2/5}{\sqrt{2}\sigma_{\overline{C}}}\right),
\end{aligned}
	\end{equation}
where 
\begin{eqnarray}
h_0(2/5)=\frac{1}{2}
{\rm Erfc}\left(\frac{1}{\sqrt{2}}\frac{2/5}{\sigma_{\overline{C}}}\right),
\end{eqnarray}
$\sigma_{\overline{C}}$ is the variance,  $H_n$ are Hermite polynomials and we  have defined in Eq. (\ref{11}) the parameters $c_n$ as 
\begin{eqnarray}
    c_n=
  \sum_{\hat{p}[n]}\,\,\,\,\prod_{\substack{p_1m_1+\cdots p_rm_r=n\\p_i\geq 0,m_i\geq 3}}
\frac{n!}{m_1!\cdots m_r! \, p_1!\cdots p_r!}\,\langle\overline{C}^{m_1}\rangle^{p_1}\cdots \langle\overline{C}^{m_r}\rangle^{p_r}, \label{110}
\end{eqnarray}
where  $\hat{p}[n]$ denotes the partitions of the integer $n$ into numbers $m_i\geq 3$. 

Given the statistics of the curvature perturbation, one can calculate the abundance of PBHs using the volume average of the linear compaction function, relying solely on superhorizon quantities. Generally, determining the statistics of the curvature perturbation can be challenging  and computing the connected cumulants is highly non-trivial. We left this task for future investigation.


\vskip 0.5cm
\centerline{\bf Acknowledgements}
\vskip 0.2cm
\noindent
We thank V. De Luca  and G. Franciolini for useful comments on the draft.
A.I. and A.R.  acknowledge support from the  Swiss National Science Foundation (project number CRSII5\_213497).
A.K. is supported by the PEVE-2020 NTUA programme for basic research with project number 65228100.
D.~P. and A.R. are supported by the Boninchi Foundation for the project ``PBHs in the Era of GW Astronomy''.

\noindent


\end{document}